\documentclass[reprint,aps,pra,superscriptaddress,showpacs,footinbib,longbibliography,amsmath,amssymb]{revtex4-2}
\usepackage{graphicx}
\usepackage{bm}

\usepackage{color}
\usepackage[colorlinks,bookmarks=false,citecolor=darkblue,linkcolor=red,urlcolor=blue]{hyperref}
\definecolor{darkblue}{rgb}{0,0.02,0.45}
\newcommand{\cred}{}
\newcommand{\resub}{}

\graphicspath{{../figures/}}

\begin{document}
\title{Magnetic hexamers interacting in layers in the (Na,K)$_2$Cu$_3$O(SO$_4)_3$ minerals}

\author{Diana O. Nekrasova}
\affiliation{Unit\'e de Catalyse et Chimie du Solide, Universit\'e Lille, 59000 Lille, France}
\affiliation{Department of Crystallography, St. Petersburg State University, 199034 Saint Petersburg, Russia}

\author{Alexander A. Tsirlin}
\email{altsirlin@gmail.com}
\affiliation{Experimental Physics VI, Center for Electronic Correlations and Magnetism, University of Augsburg, 86159 Augsburg, Germany}

\author{Marie Colmont}
\affiliation{Unit\'e de Catalyse et Chimie du Solide, Universit\'e Lille, 59000 Lille, France}

\author{Oleg Siidra}
\affiliation{Department of Crystallography, St. Petersburg State University, 199034 Saint Petersburg, Russia}
\affiliation{Kola Science Center, Russian Academy of Sciences, 184200 Apatity, Russia}

\author{Herv\'e Vezin}
\affiliation{Laboratoire de Spectrochimie Infrarouge et Raman (LASIR), UMR 8516, Universit\'e Lille, 59655 Villeneuve d’ASCQ, France}

\author{Olivier Mentr\'e}
\email{olivier.mentre@univ-lille.fr}
\affiliation{Unit\'e de Catalyse et Chimie du Solide, Universit\'e Lille, 59000 Lille, France}

\begin{abstract}
Magnetic properties and underlying magnetic models of the synthetic A$_2$Cu$_3$O(SO$_4)_3$ fedotovite (A = K) and puninite (A = Na) minerals, as well as the mixed euchlorine-type NaKCu$_3$O(SO$_4)_3$ are reported. We show that all three compounds contain magnetic Cu$_6$ hexamer units, which at temperatures below about 100\,K act as single spin-1 entities. Weak interactions between these magnetic molecules lead to long-range order below $T_N=3.4$\,K (A = Na), 4.7\,K (A = NaK), and about 3.0\,K (A = K). The formation of the magnetic order is elucidated by \textit{ab initio} calculations that reveal two-dimensional inter-hexamer interactions within crystallographic $bc$ planes. This model indicates the presence of a weakly distorted square lattice of $S=1$ magnetic ions and challenges the earlier description of the A$_2$Cu$_3$O(SO$_4)_3$ minerals in terms of Haldane spin chains. 
\end{abstract}

\maketitle

\section{Introduction}
Interesting quantum phenomena can be achieved in both finite clusters~\cite{coronado2020} and periodic lattices~\cite{knolle2019} of magnetic ions. Some of the known materials straddle the border between the two, because they feature small magnetic clusters (magnetic molecules) that interact and eventually form a periodic lattice, where each cluster acts as a single magnetic site. The interplay of different dimensionalities and energy scales associated with interactions within and between such clusters can lead to unusual physics. For example, Cu$_2$OSeO$_3$ with a chiral lattice built by the Cu$_4$ tetrahedra hosts several skyrmion phases~\cite{chacon2018,qian2018} tunable by electric field~\cite{seki2012,white2014}, reveals non-reciprocal propagation of phonons~\cite{nomura2019}, and shows topological magnon states~\cite{zhang2020}. Developing microscopic magnetic models of such complex systems is far from trivial, though. In Cu$_2$OSeO$_3$, the presence of Cu$_4$ tetrahedra acting as single magnetic units was hardly appreciated, until pointed out by \textit{ab initio} calculations~\cite{janson2014} and further confirmed spectroscopically~\cite{ozerov2014,portnichenko2016,tucker2016}. 

\begin{figure}
\includegraphics[width=6cm]{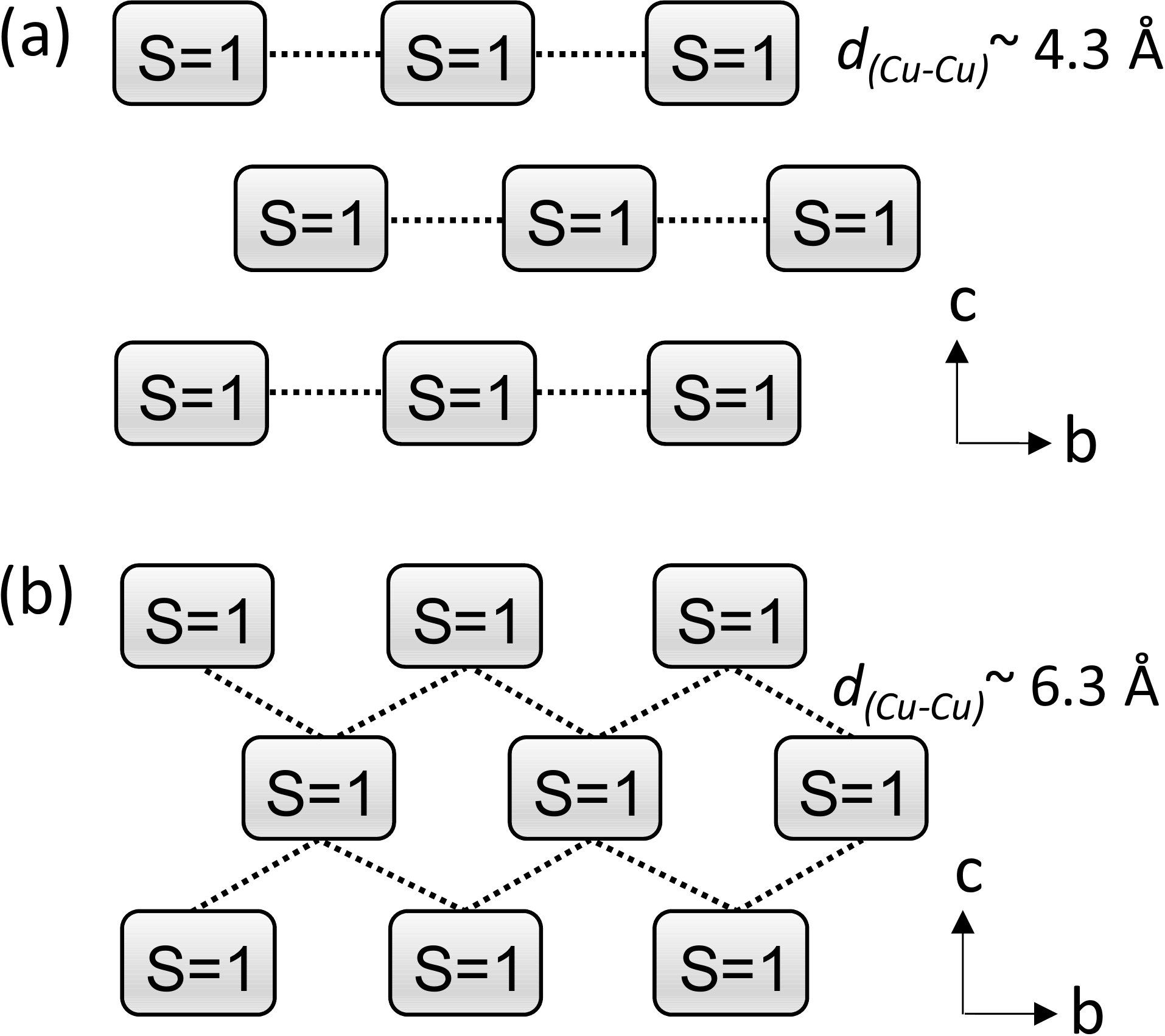}
\caption{\label{fig:intro}
Possible magnetic models of A$_2$Cu$_3$O(SO$_4)_3$: (a) Haldane chains postulated in Ref.~\onlinecite{fujihala2018}; (b) $S=1$ square lattice established in the present work. Each $S=1$ is the Cu$_6$ hexamer unit shown in detail in Fig.~\ref{fig:structure}.
}
\end{figure}

Here, we focus on a less explored family of Cu$^{2+}$-based quantum magnets A$_2$Cu$_3$O(SO$_4)_3$ realized in the minerals fedotovite (A = K)~\cite{starova1991,lander2017} and puninite (A = Na)~\cite{siidra2017}. In these compounds, the presence of Cu$_6$ hexamer units with the $S=1$ ground state is already well-established by neutron spectroscopy~\cite{fujihala2018,furrer2018,furrer2020}, but weaker interactions between the hexamers remain controversial. The linear arrangement of the hexamers along the crystallographic $b$ direction (Fig.~\ref{fig:intro}a) led to an idea~\cite{fujihala2018} that their $S=1$ units may form Haldane chains~\cite{haldane1983,affleck1989} and develop a gapped ground state without long-range magnetic order. Small spin gap of about 0.6\,meV observed by inelastic neutron scattering corroborated this interpretation~\cite{fujihala2018,furrer2018}, although the simultaneous presence of magnetic Bragg peaks and even a weak thermodynamic anomaly at 3.0\,K in the A = K compound~\cite{hase2019} suggested that long-range magnetic order sets in. 
\resub{Little is known about the magnetism of Na$_2$Cu$_3$O(SO$_4)_3$ except the fact that it also contains $S=1$ hexamer units and develops a spin gap of about 0.6\,meV at low temperatures~\cite{furrer2018}. Moreover, no attempts to evaluate the strength and dimensionality of magnetic interactions between the hexamers were performed.} 

Here, we address these pending issues by studying thermodynamic properties of, and magnetic interactions in fedotovite (A = K), puninite (A = Na), and the mixed euchlorine-type NaKCu$_3$O(SO$_4)_3$~\cite{scordari1990}. \resub{From thermodynamic measurements, we show the formation of magnetic order and evaluate N\'eel temperatures as well as energy scales of the inter-hexamer interactions.} Contrarily to the earlier picture, our data reveal that magnetic hexamers interact along diagonals in the $bc$ plane (Fig.~\ref{fig:intro}b) and not directly along $b$, where Haldane chains would form. Our revised model explains \cred{the formation of long-range magnetic order}, caused by the two-dimensional (2D) coupling geometry along with the single-ion anisotropy of the $S=1$ units of individual hexamers. We conclude that none of the compounds reveal one-dimensional coupling geometry implied by the Haldane model, and long-range magnetic order sets in at temperatures comparable to the energy scale of magnetic couplings. All these observations leave little room for the Haldane physics in this family of compounds \resub{and set the scene for investigating their spin dynamics}.

\section{Methods}
\textbf{Synthesis.} Single-phase sulfate materials were prepared by a solid-state reaction from a stoichiometric mixture of the anhydrous precursors A$_2$(SO$_4$) (A = Na, K), CuSO$_4$, and CuO taken in the 1:2:1 ratio. The mixtures were ground in an agate mortar, loaded into gold plates, kept at 560\,$^{\circ}$C for 3 hours in air, and subsequently cooled for 9 hours to room temperature. The resulting solid products are inhomogeneous in texture but contain single crystals (typical crystal size $0.15\times 0.15\times 0.10$\,mm). The results of our single crystal x-ray diffraction (XRD) refinements are given in the Supplemental Material~\cite{supplement}.

\begin{figure*}
\includegraphics[width=\textwidth]{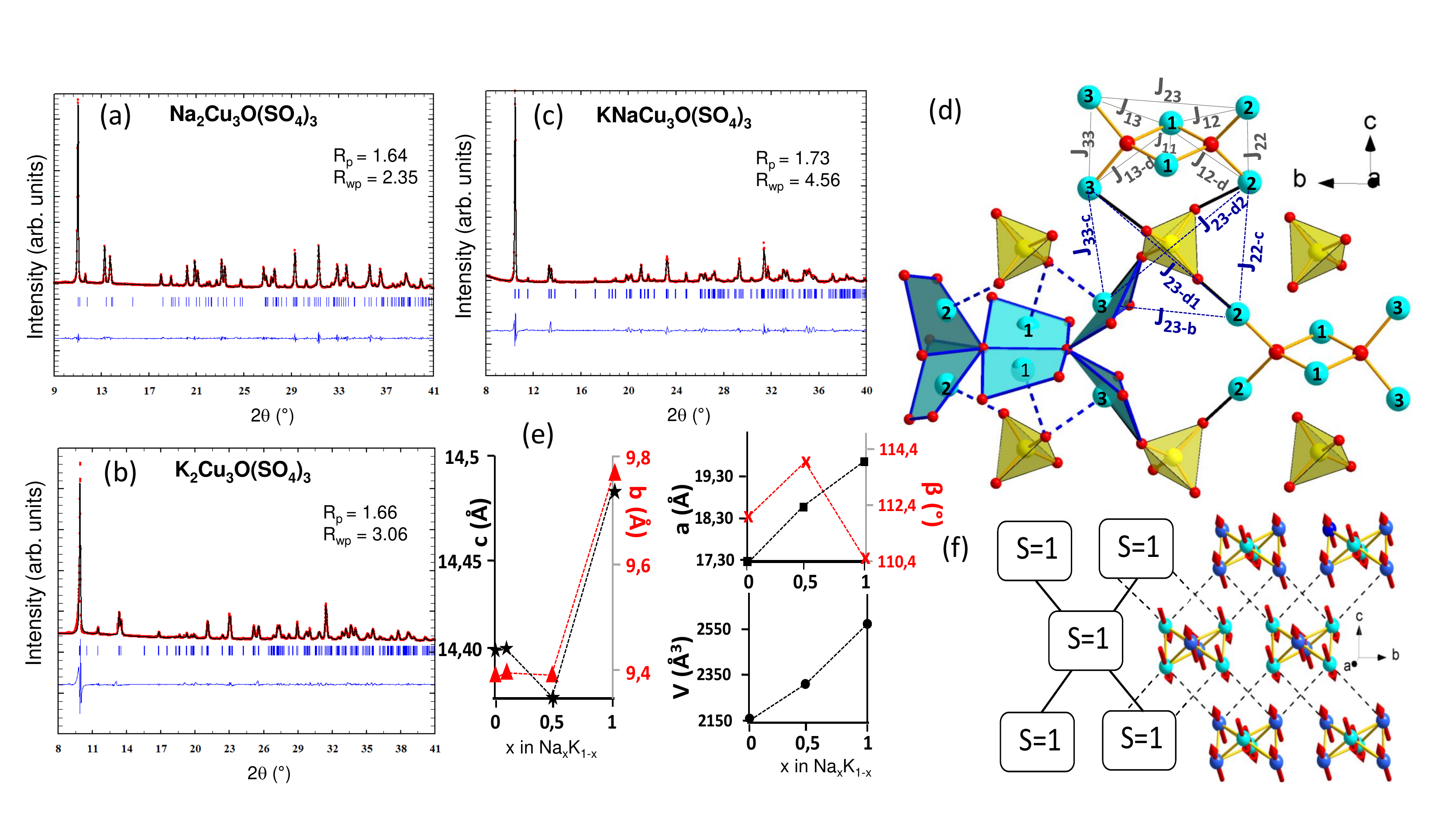}
\caption{\label{fig:structure}
(a,b,c) Profile-matching refinement and refinement residuals for A$_2$Cu$_3$O(SO$_4)_3$.(d) Crystal structure and main exchange couplings (intra-hexamer: blue lines, inter-hexamer: grey lines); blue color -- Cu$^{2+}$, red -- O$^{-2}$, yellow -- S$^{6+}$. The longest apical Cu--O4+1 bonds are shown by the dashed blue lines. (e) Lattice parameters of the Na$_{1-x}$K$_x$ solid solution. (f) Two-dimensional square lattice of the $S=1$ hexamer units with the magnetic structure of Ref.~\onlinecite{hase2019} assuming the $(u,0,u)$ spin components. Dotted lines denote the leading magnetic interactions $J_{23-d1}$ and $J_{23-d2}$ between the hexamers.
}
\end{figure*}

\textbf{Powder X-ray diffraction.} Powder XRD patterns were collected at room temperature in the $2\theta$ range of $10-110^{\circ}$ using the Bruker D8 diffractometer. The Profile-matching refinements were carried out using JANA2006~\cite{jana2006}. The background was fitted using Chebyshev polynomial function, and the peak shapes were described by a Pseudo-Voigt function. The results are shown Fig.~\ref{fig:structure} and confirm single-phase nature of our samples. The samples are sensitive to air moisture and have been handled in Ar-filled glove box, although short contact with air could not be avoided when preparing the samples for thermodynamic measurements.

\textbf{Thermodynamic properties.} Magnetization and heat capacity were studied on powder samples using the PPMS Dynacool (9\,T) from Quantum Design. For temperature-dependent magnetization measurements both zero-field-cooling (ZFC) and field-cooling (FC) protocols were used. Magnetization versus field was measured at 2\,K and 300\,K. Specific heat was measured on pressed pellets from 1.9\,K to 300\,K in zero field. 

\textbf{Electron spin resonance.} X-band electron paramagnetic resonance (EPR) experiments were carried out with a Bruker ELEXYS E580E spectrometer. Microwave power and modulation amplitude were 1\,mW and 5\,G, respectively. The spectra were recorded between 300 and 4\,K using helium ITC503 Oxford temperature control.

\textbf{\textit{Ab initio} calculations.} Exchange couplings between the Cu$^{2+}$ ions were obtained by density-functional-theory (DFT) band-structure calculations performed in the \texttt{FPLO} code~\cite{fplo} with the Perdew-Burke-Ernzerhof flavor of the exchange-correlation potential~\cite{pbe96}. Mapping procedure~\cite{xiang2011} was used to calculate exchange parameters $J_{ij}$ of the spin Hamiltonian,
\begin{equation}
 \mathcal H_{\rm Cu}=\sum_{\langle ij\rangle}J_{ij}\,\mathbf s_i\mathbf s_j
\label{eq:ham}\end{equation}
where the summation is over lattice bonds $\langle ij\rangle$, and $s=\frac12$ for individual Cu$^{2+}$ ions. Correlation effects in the Cu $3d$ shell were treated on the mean-field level using the DFT+$U$ procedure with the on-site Coulomb repulsion $U_d=9.5$\,eV, Hund's coupling $J_d=1$\,eV, and double-counting correction in the atomic limit~\cite{nath2013,nath2015}.

\textbf{Quantum Monte-Carlo simulations.} Magnetic susceptibility and magnetization were calculated using the \texttt{loop}~\cite{loop} and \texttt{dirloop\_sse}~\cite{dirloop} algorithms of the \texttt{ALPS} simulation package~\cite{alps} on finite lattices with periodic boundary conditions and up to $L=48$ or $L=12\times 12$ sites for the 1D and 2D interaction geometries, respectively. The Heisenberg spin Hamiltonian akin to that of Eq.~\eqref{eq:ham} was augmented with a single-ion anisotropy term
\begin{equation}
 \mathcal H_{\rm hexamer}=\sum_{\langle ij\rangle}\mathfrak J\,\mathbf S_i\mathbf S_j+\sum_i \mathfrak D(S_i^z)^2
\label{eq:ham2}\end{equation}
for $S=1$ of the Cu$_6$ hexamer, where $\mathfrak J$ labels interactions between these $S=1$ units, as opposed to the interactions $J_{ij}$ between the individual Cu$^{2+}$ spins, and $\mathfrak D$ stands for the single-ion anisotropy. Magnetic ordering temperatures were determined by analyzing spin stiffness for different lattice size~\cite{sandvik1998}. 

\begin{table}
\caption{\label{tab:moments}
Parameters of the Curie-Weiss fits $\chi_m=C/(T-\theta)$: effective moments $\mu_{\rm eff}$ (in\,$\mu_B$) and Curie-Weiss temperatures $\theta$ (in\,K) for the low-temperature ($T<100$\,K) and high-temperature ($T>150$\,K) regions. The $\mu_{\rm eff}$ values are normalized per Cu for the high-$T$ part and per Cu$_6$ hexamer for the low-$T$ part.
}
\begin{ruledtabular}
\begin{tabular}{c@{\hspace{2em}}cc@{\hspace{2em}}cc@{\hspace{2em}}cc}
 & \multicolumn{2}{l}{\,\,\,A = Na} & \multicolumn{2}{l}{\,\,A = NaK} & \multicolumn{2}{l}{\quad A = K} \\
 & $\mu_{\rm eff}$ & $\theta$ & $\mu_{\rm eff}$ & $\theta$ & $\mu_{\rm eff}$ & $\theta$ \\\hline
high-$T$ & 1.93 & $-235$ & 1.73 & $-187$ & 1.80 & $-216$ \\
low-$T$  & 2.84 & $-7.3$ & 2.71 & $-8.4$ & 2.80 & $-10.7$ \\
\end{tabular}
\end{ruledtabular}
\end{table}

\section{Results and Discussion}

\subsection{Individual hexamers}

The Cu$_6$ hexamers are easily recognized in the A$_2$Cu$_3$O(SO$_4)_3$ structure as units built by six CuO$_4$ plaquettes sharing edges and/or corners. The Cu--Cu separations are under 3.5\,\r A within the hexamer and above 4.3\,\r A between the hexamers, suggesting that leading magnetic interactions may be restricted to individual Cu$_6$ units (Fig.~\ref{fig:structure}d). Indeed, inelastic neutron scattering on K$_2$Cu$_3$O(SO$_4)_3$ and Na$_2$Cu$_3$O(SO$_4)_3$ revealed several sharp molecular-like excitations that had been used to identify the $S=1$ ground state and gauge, albeit with some ambiguity, magnetic interactions within the hexamer~\cite{furrer2018,furrer2020}. 

\begin{figure*}
\includegraphics[width=\textwidth]{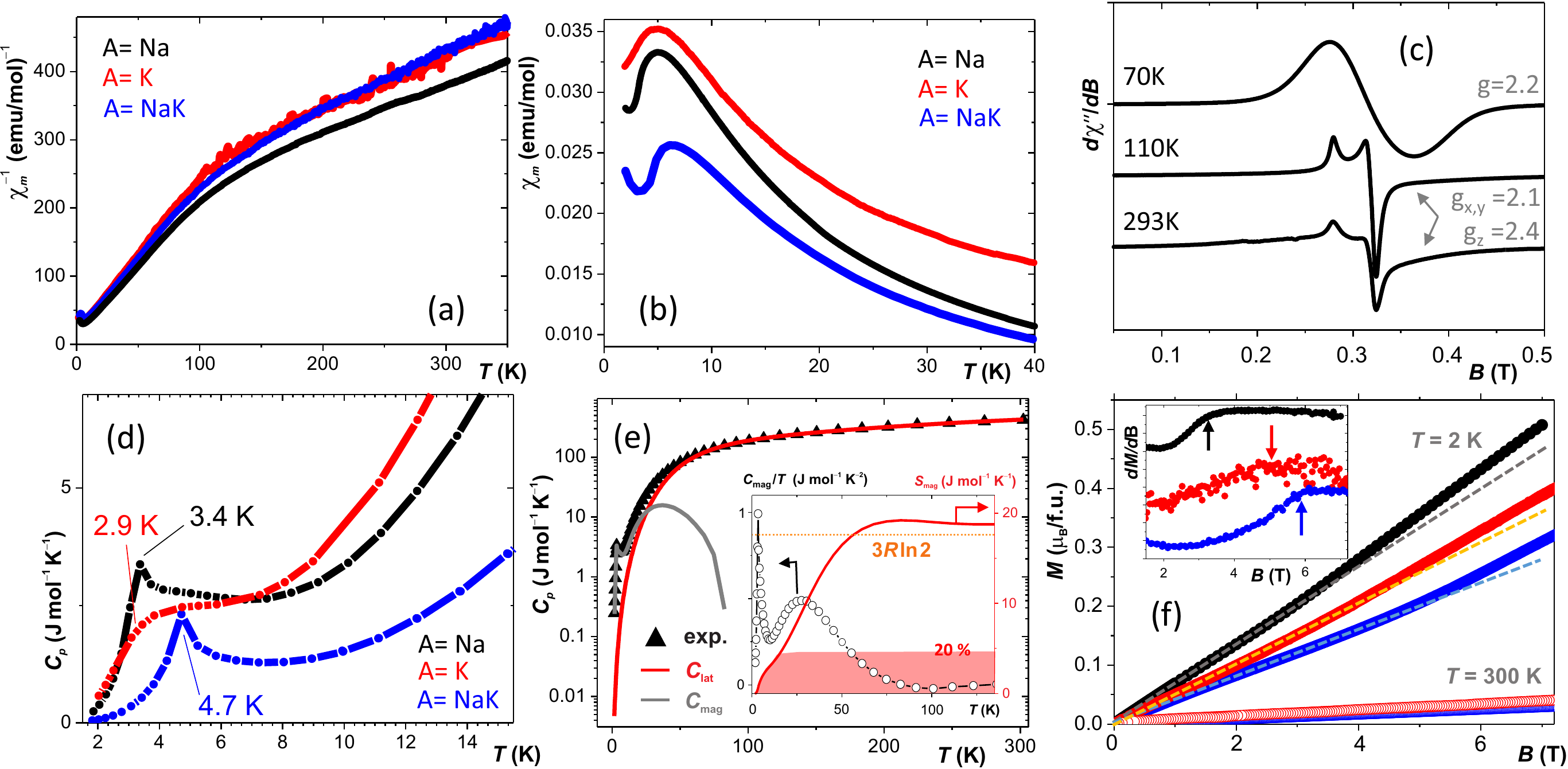}
\caption{\label{fig:properties}
(a) Inverse molar susceptibility ($\chi_m$) measured at 0.1\,T for the A$_2$Cu$_3$O(SO$_4)_3$ series. The data for the A = K sample are corrected by subtracting a temperature-independent background contribution. (b) Low-temperature magnetic susceptibility showing a bump typical of low-dimensional magnetism. (c) ESR spectra for Na$_2$Cu$_3$O(SO$_4)_3$ taken at several temperatures. (d) Temperature dependence of the specific heat at low temperatures, with $\lambda$-type peaks clearly seen for A = Na and A = NaK. A similar feature is expected for A = K around 3.0\,K~\cite{hase2019}. (e) Phonon and magnetic contributions to the specific heat of Na$_2$Cu$_3$O(SO$_4)_3$. The inset shows magnetic entropy $S_{\rm mag}$ obtained by integrating $C_{\rm mag}/T$, with about 20\,\% of the total entropy recovered within the lower-temperature peak. (f) Raw $M(H)$ data at 2\,K and 300\,K with the spin-flop transition highlighted via $dM/dH$ in the inset. 
}
\end{figure*}

From thermodynamic perspective, the presence of finite magnetic clusters can be inferred from two linear regimes in the inverse susceptibility (Fig.~\ref{fig:properties}a), where the linear part above 150\,K corresponds to individual Cu$^{2+}$ ions with the paramagnetic effective moment of $\mu_{\rm eff}=1.73-1.93$\,$\mu_B$ per Cu$^{2+}$ (Table~\ref{tab:moments}) closely matching $1.73$\,$\mu_B$ expected for spin-$\frac12$. On the other hand, the linear part below 100\,K with $\mu_{\rm eff}=2.71-2.84$\,$\mu_B$ per hexamer is compatible with 2.83\,$\mu_B$ expected for spin-1. The crossover around 120\,K (thermal energy of 10.3\,meV) is also consistent with 12.6\,meV and 13.5\,meV as the energy separation between the ground state and first excited state in K$_2$Cu$_3$O(SO$_4)_3$ and Na$_2$Cu$_3$O(SO$_4)_3$, respectively~\cite{furrer2020}. 

As the magnetism of Na$_2$Cu$_3$O(SO$_4)_3$ has not been characterized in detail, we also performed ESR and specific heat measurements on that sample. At room temperature, the ESR spectrum is well described by the powder average of two lines with $g_{\|}=2.4$ and $g_{\perp}=2.1$ (Fig.~\ref{fig:properties}c), resulting in $g_{\rm av}=2.20$ and $\mu_{\rm eff}=g\mu_B\sqrt{S(S+1)}=1.90$\,$\mu_B$ in excellent agreement with $1.93$\,$\mu_B$ obtained from the susceptibility data (Table~\ref{tab:moments}). Below 110\,K, the ESR line broadens, indicating the crossover to the collective behavior of Cu$^{2+}$ spins within the Cu$_6$ hexamer unit.

To extract magnetic specific heat $C_{\rm mag}$, we first fitted the high-temperature part with
\begin{equation}
 C_p=9R\sum_{i=1}^2 c_i\left(\frac{T}{\theta_{Di}}\right)^3\int\limits_0^{\theta_{Di}/T}\frac{x^4e^x}{(e^x-1)^2}\,dx
\end{equation}
to determine the phonon contribution. Using $c_1=11(1)$, $\theta_{D1}=310(20)$\,K, $c_2=18(1)$, $\theta_{D2}=1502(50)$\,K, we obtain magnetic specific heat that, upon integrating $C_{\rm mag}/T$, yields full magnetic entropy of $3R\ln 2$ above 100\,K (Fig.~\ref{fig:properties}e). At low temperatures, the magnetic entropy is released in two steps corresponding to two maxima in $C_{\rm mag}/T$, the lower one containing about 20\,\% of the total entropy and indicating collective behavior of the hexamers that are responsible for $\frac16$ (16.7\,\%) of spin degrees of freedom.

\begin{table*}
\caption{\label{tab:intra}
Magnetic interactions within the Cu$_6$ hexamers: the Cu--Cu distances $d_i$ (in\,\r A), Cu--O--Cu bond angles $\alpha_i$ (in\,deg), and exchange couplings $J_i$ (in\,K) obtained \textit{ab initio} using DFT+$U$ (this work) or by fitting magnetic excitation energies with a seven-parameter model (Ref.~\onlinecite{furrer2020}, A = Na and A = K only). 
}
\begin{ruledtabular}
\begin{tabular}{c@{\hspace{3em}}cccc@{\hspace{3em}}cccc@{\hspace{3em}}ccc}
 & \multicolumn{4}{l}{\qquad\qquad\qquad A = Na} & \multicolumn{4}{l}{\qquad\qquad\qquad A = K} & \multicolumn{3}{c}{A = NaK} \\

 & $d$ & $\alpha$ & $J_{\rm DFT+U}$ & $J_{\rm neutron}$ & $d$ & $\alpha$ & $J_{\rm DFT+U}$ & $J_{\rm neutron}$ & $d$ & $\alpha$ & $J_{\rm DFT+U}$ \\\hline

 $J_{11}$   & 2.820 & 92.2/93.3 & $-159$ & $-268$ & 2.818 & 91.4/92.4 & $-157$ & $-289$ & 2.807 & 91.4/91.8 & $-164$ \\
 $J_{12}$   & 3.394 & 121.8     & 171    & 194    & 3.401 & 122.3     & 171    & 184    & 3.391 & 122.2     & 170   \\
 $J_{12-d}$ & 3.180 & 109.8     & 170    & 124    & 3.190 & 110.2     & 153    & 121    & 3.167 & 109.6     & 141   \\
 $J_{13}$   & 3.111 & 107.2     & 118    & 110    & 3.117 & 107.9     & 139    & 107    & 3.135 & 108.3     & 145   \\
 \smallskip
 $J_{13-d}$ & 3.419 & 124.3     & 230    & 194    & 3.416 & 124.3     & 198    & 184    & 3.393 & 122.1     & 190   \\ 
 $J_{22}$   & 3.014 & 102.7     & 26     & $-51$  & 2.977 & 102.0     &  42    &  58    & 2.999 & 103.1     & 36    \\
 $J_{33}$   & 3.007 & 102.5     & $-2$   & $-30$  & 2.977 & 102.0     &  19    &  37    & 3.019 & 104.2     & 29    \\
 $J_{23}$   & 5.133 & --        & 16     & --     & 5.165 & --        &  14    &  --    & 5.124 &  --       & 14    \\ 
\end{tabular}
\end{ruledtabular}

\end{table*}

On the microscopic level, the Cu$_6$ hexamer can be seen as two condensed Cu$_4$ tetrahedra, each centered by a single oxygen atom. Such oxo-centered OCu$_4$ tetrahedra are very common in Cu$^{2+}$ compounds~\cite{krivovichev2013}, including many of the copper minerals, and usually lead to magnetic frustration, because antiferromagnetic interactions compete on each face of the tetrahedron. This is not the case in A$_2$Cu$_3$O(SO$_4)_3$, though. Inelastic neutron scattering data are consistent with the combination of ferromagnetic (FM) and antiferromagnetic (AFM) interactions that eventually release the frustration within the hexamer unit~\cite{fujihala2018,furrer2018}. 

Our \textit{ab initio} results support this scenario and indicate FM $J_{11}$ along with AFM $J_{12}$, $J_{12-d}$, $J_{13}$, and $J_{13-d}$. Absolute values of leading exchange couplings are in favorable agreement with the earlier estimates from neutron spectroscopy (Table~\ref{tab:intra}). \cred{Remaining discrepancies may be caused by systematic errors involved in \textit{ab initio} calculations, especially for $J_{11}$ with the shortest Cu--Cu distance that leads to an interplay of potential and kinetic exchange and is usually most difficult to estimate from \textit{ab initio}~\cite{badrtdinov2019}.} We also note that previous neutron studies assumed the FM sign of $J_{22}$ and $J_{33}$ and neglected $J_{23}$, while in \textit{ab initio} all these couplings are AFM and cause a weak frustration of the hexamer unit.

The hierarchy of intra-hexamer interactions follows the Cu--O--Cu angles (Table~\ref{tab:intra}), as expected from Goodenough-Kanamori-Anderson rules. The nearly $90^{\circ}$ bond angles render $J_{11}$ ferromagnetic. In contrast, the angles above $100^{\circ}$ give rise to AFM interactions. Weaker $J_{22}$ and $J_{33}$ feature the bond angles of $102-104^{\circ}$, while stronger interactions $J_{12}$, $J_{12-d}$, $J_{13}$, and $J_{13-d}$ are characterized by the bond angles between $107^{\circ}$ and $125^{\circ}$. These simple geometrical arguments do not explain why $J_{12}$ and $J_{12-d}$ or $J_{13}$ and $J_{13-d}$ are of similar size despite more than $10^{\circ}$ difference in their bond angles, but here the Cu--Cu distances probably play a role, with the shorter distances of $J_{12-d}$ and $J_{13-d}$ facilitating AFM exchange via the direct $d-d$ hopping~\cite{badrtdinov2019}. 

It is also worth noting that the mixed NaKCu$_3$O(SO$_4)_3$ compound is not a simple intermediate between the limiting cases of A = Na and A = K (Table~\ref{tab:intra}). Indeed, the lattice parameters change non-monotonically upon the Na-K substitution (Fig.~\ref{fig:structure}e), reflecting the fact that K preferentially occupies the Na1 site of the Na$_2$Cu$_3$O(SO$_4)_3$ structure. NaKCu$_3$O(SO$_4)_3$ shows the ideal site order of Na and K~\cite{scordari1990}.


\subsection{Interactions between the hexamers}

Below 100\,K, the Cu$_6$ hexamers act as single $S=1$ units that, according to the previous studies~\cite{fujihala2018,furrer2018}, should build Haldane chains and develop a spin gap, while showing no magnetic order. This interpretation would be consistent with broad susceptibility maxima observed at $6-7$\,K in all three compounds (Fig.~\ref{fig:properties}b). On the other hand, the A = Na and A = NaK compounds clearly show long-range magnetic order with $T_N=3.4$\,K and 4.7\,K, respectively, as evidenced by the weak kinks in $\chi_m(T)$ and sharp $\lambda$-type anomalies in the specific heat (Fig.~\ref{fig:properties}d). No clear transition anomaly could be seen in the case of A = K, probably because this sample is more sensitive to air moisture and may have slightly deteriorated upon transferring to the PPMS. However, earlier thermodynamic measurements as well as neutron diffraction data~\cite{hase2019} indicate the formation of magnetic long-range order also in this compound below about 3\,K.

\begin{figure}
\includegraphics{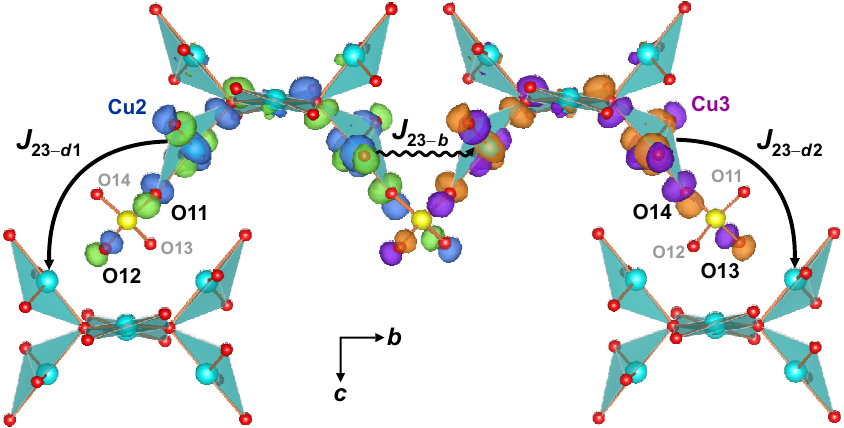}
\caption{\label{fig:wannier}
Superexchange mechanism for the inter-hexamer couplings $J_{23-d1}$ and $J_{23-d2}$. Magnetic orbital of Cu$^{2+}$ contains a sizable contribution of the second-neighbor oxygen (O12 for Cu2 and O13 for Cu3) that is proximate to the Cu atom of the adjacent hexamer. In contrast, no interaction $J_{23-b}$ occurs, because magnetic orbitals of the interacting Cu$^{2+}$ ions contain contributions from different oxygen atoms and do not overlap. The figure was prepared using the \texttt{VESTA} software~\cite{vesta}.
}
\end{figure}

\begin{figure*}
\includegraphics{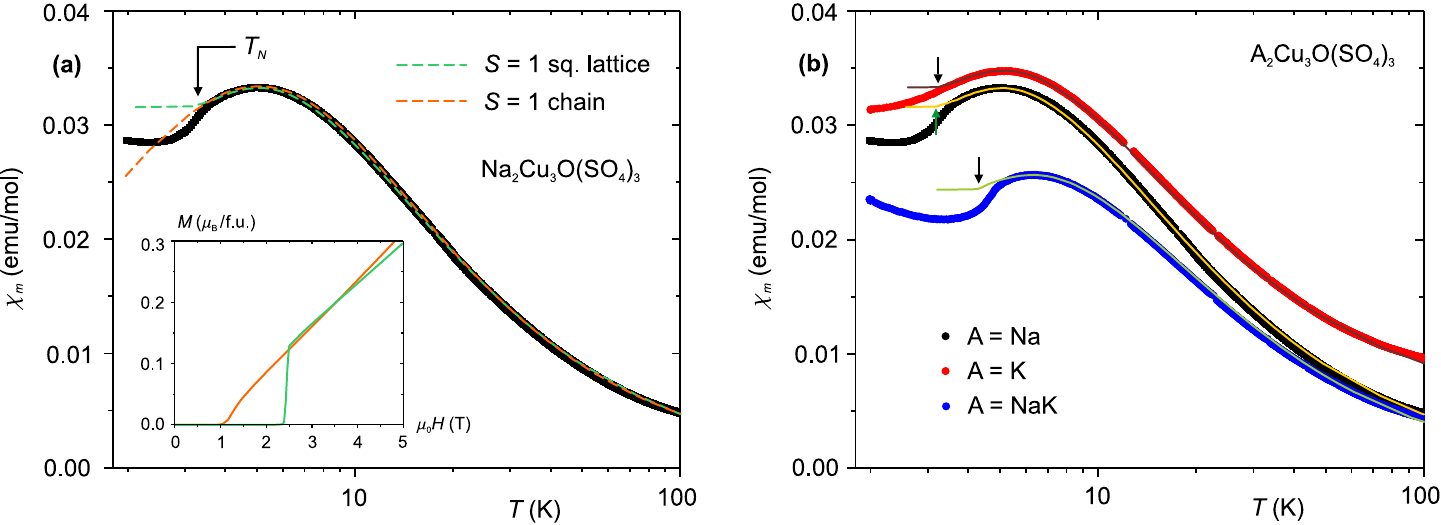}
\caption{\label{fig:fits}
Fits to the low-temperature magnetic susceptibility using the $S=1$ model of Eq.~\eqref{eq:ham2}. (a) Comparison of the 1D (Haldane chain, $\mathfrak D=0$) and 2D (square lattice, $\mathfrak D\neq 0$) models for Na$_2$Cu$_3$O(SO$_4)_3$; the inset shows magnetization curves calculated at low temperatures to highlight the larger gap expected in the 2D case. (b) Fits for all three compounds with the 2D model; the arrows show transition temperatures determined from quantum Monte-Carlo simulations.
}
\end{figure*}

From the \textit{ab initio} perspective, the coupling $J_{23-b}$ forming the presumed Haldane chains is zero within the accuracy of our method (Table~\ref{tab:inter}). On the other hand, we find sizable couplings $J_{23-d1}$ and $J_{23-d2}$ that connect the hexamers in the $bc$ plane. \cred{This coupling topology is verified by neutron diffraction data for the K compound that demonstrated the propagation vector $\mathbf k=0$~\cite{hase2019}. In contrast, Haldane chains with $J_{23-b}>J_{23-d1},J_{23-d2}$ would necessarily cause antiferromagnetic order along $b$ and the doubling of the magnetic unit cell along this direction (Fig.~\ref{fig:intro}a). The resulting propagation vector of $\mathbf k=(0,\frac12,0)$ is incompatible with the experimental observation of magnetic Bragg peaks at integer positions~\cite{hase2019}.}

\cred{The magnetic structure stabilized by $J_{23-d1}$ and $J_{23-d2}$ fully respects the experimental propagation vector $\mathbf k=0$ and the magnetic Shubnikov group $C2'/c$ inferred from the neutron diffraction data. According to Ref.~\onlinecite{hase2019}, the $(u,v,w)$ spin components are dominated by the $u$ and $w$ terms. In Fig.~\ref{fig:structure}f, we arbitrarily choose $u=w$ for better visualization and demonstrate that the magnetic order associated with $C2'/c$ complies with the antiferromagnetic nature of $J_{23-d1}$ and $J_{23-d2}$, as well as with all other couplings obtained from our DFT calculations (Tables~\ref{tab:intra} and~\ref{tab:inter}).} 

\begin{table}
\caption{\label{tab:inter}
Magnetic interactions between the hexamers: the Cu--Cu distances $d_i$ (in\,\r A) and exchange couplings $J_i$ (in\,K) obtained from DFT+$U$. The last two lines contain experimental values of the coupling $\mathfrak J$ (in\,K) and single-ion anisotropy $\mathfrak D$ (in\,K) extracted from fits to the experimental susceptibility data using the $S=1$ square-lattice model and $g=2.04$ for all three compounds (Fig.~\ref{fig:fits}).
}
\begin{ruledtabular}
\begin{tabular}{c@{\hspace{2em}}cc@{\hspace{2em}}cc@{\hspace{2em}}cc}
 & \multicolumn{2}{l}{\,\,\,A = Na} & \multicolumn{2}{l}{\quad A = K} & \multicolumn{2}{l}{A = NaK} \\
 & $d_i$ & $J_i$ & $d_i$ & $J_i$ & $d_i$ & $J_i$ \\\hline
 $J_{23-b}$  & 4.344 & 0 & 4.444 & 0 & 4.313 & 0 \\ 
 $J_{23-d1}$ & 6.349 & 7 & 6.333 & 9 & 6.270 & 9 \\
 $J_{23-d2}$ & 6.332 & 9 & 6.378 & 9 & 6.375 & 11 \\\hline
 $\mathfrak J$ & \multicolumn{2}{l}{\quad\,\,\,\,2.45} & \multicolumn{2}{l}{\quad\,\,\,\, 2.48} & \multicolumn{2}{l}{\quad\,\,\,\, 3.05} \\
 $\mathfrak D$ & \multicolumn{2}{l}{\quad\,\,\,\,0.74} & \multicolumn{2}{l}{\quad\,\,\,\,\,\,0.74} & \multicolumn{2}{l}{\quad\,\,\,\,\,\,0.92} \\
\end{tabular}
\end{ruledtabular}
\end{table}

The $J_{23-b}\ll J_{23-d1}$, $J_{23-d2}$ regime may look counter-intuitive at first glance, given the much shorter Cu--Cu distance for the former coupling. A closer look at the superexchange pathways (Fig.~\ref{fig:wannier}) reveals that the magnetic orbital of Cu$^{2+}$ contains a sizable contribution from the second-neighbor oxygen atom, which is proximate to the Cu atom of the adjacent hexamer. This peculiarity of the magnetic orbital facilitates $J_{23-d1}$ and $J_{23-d2}$ and at the same time precludes the coupling $J_{23-b}$, because in this case magnetic orbitals of the interacting Cu$^{2+}$ ions include contributions from different oxygen atoms and do not overlap (Fig.~\ref{fig:wannier}). 

Interestingly, the tail of the magnetic orbital includes only one second-neighbor oxygen atom of the SO$_4$ tetrahedron. This atom is chosen by the Cu--O--O angle, which in the case of Cu2 in Na$_2$Cu$_3$O(SO$_4)_3$ is $168.4^{\circ}$ for O12, $110.2^{\circ}$ for O13, and $109.0^{\circ}$ for O14, hence only O12 gives a sizable contribution. Another important observation is that all Cu atoms feature a very distorted local environment that can be better described as a CuO$_{4+1}$ square pyramid (Fig.~\ref{fig:structure}d). The magnetic orbital is of $d_{x^2-y^2}$ nature and mostly restricted to the basal plane of the pyramid. Nevertheless, an admixture of the $d_{3z^2-r^2}$ orbital exists and allows the participation of the apical oxygen atom in the superexchange. This apical oxygen atom features a relatively short separation to copper (2.19\,\r A for Cu2-O13 and 2.32\,\r A for Cu3-O12), whereas the longer apical contacts of Cu1 (2.56\,\r A for Cu1-O12) do not play any significant role in the superexchange, because they are well above the sum of the van der Waals radii. The inclusion of the apical oxygen atom allows to analyze the interactions between the hexamers in terms of Cu--O--O--Cu dihedral angles, which are $75.9^{\circ}$ ($J_{23-b}$), $18.4^{\circ}$ ($J_{23-d1}$), and $17.7^{\circ}$ ($J_{23-d2}$), thus disfavoring $J_{23-b}$ and favoring the two other couplings.

Other possible inter-hexamer couplings were analyzed by calculating $J_{ij}$ for all Cu--Cu pairs with interatomic distances up to 8\,\r A. Long-range couplings in the $bc$ plane are below 1\,K and thus negligible compared to $J_{23-d1}$ and $J_{23-d2}$. The couplings perpendicular to the $bc$ plane are even weaker, below 0.1\,K. We thus conclude that at low temperatures all three A$_2$Cu$_3$O(SO$_4)_3$ compounds should be well described by the model of $S=1$ ions interacting on a rectangular lattice with two nonequivalent couplings arising from the Cu--Cu couplings $J_{23-d1}$ and $J_{23-d2}$. Because these two couplings are similar in magnitude, we further consider the $S=1$ square lattice with the single coupling $\mathfrak J$ as a reasonable simplification of this model.

Our microscopic magnetic model is different from the earlier model of Haldane chain in terms of both dimensionality (2D instead of 1D) and proclivity to long-range magnetic order. Weakly coupled Haldane chains should not develop long-range order, because each chain is in the gapped state. On the other hand, the $S=1$ square lattice of magnetic ions is subject to magnetic order with a finite $T_N$ in the presence of an infinitesimally small interlayer coupling or single-ion anisotropy. By fitting magnetic susceptibility of Na$_2$Cu$_3$O(SO$_4)_3$, we conclude that the anisotropy is essential to reproduce the experimental $T_N$. Indeed, with $\mathfrak J=2.45$\,K and $T_N=3.4$\,K, in the absence of anisotropy one expects the interlayer coupling $\mathfrak J_{\perp}/J\simeq 0.1$~\cite{yasuda2005} or $\mathfrak J_{\perp}\simeq 0.25$\,K, which is several times larger than our upper estimate of 0.1\,K for the Cu--Cu couplings perpendicular to the $bc$ plane. On the other hand, by adding the finite single-ion anisotropy $\mathfrak D$ we successfully reproduce both $\chi_m(T)$ and $T_N$ in all three compounds even within the 2D model, where no interlayer coupling was included (Table~\ref{tab:inter} and Fig.~\ref{fig:fits}b).

\cred{The Haldane-chain model may be capable of reproducing the experimental susceptibility too (Fig.~\ref{fig:fits}), but the resulting $\mathfrak J_{\rm 1D}\simeq 3.9$\,K (A = Na, K) leads to only a minute spin gap $0.41\mathfrak J_{\rm 1D}\simeq 1.6$\,K (0.13\,meV), which is several times smaller than 0.6\,meV reported experimentally~\cite{fujihala2018,furrer2018}. The spin gap of the A$_2$Cu$_3$O(SO$_4)_3$ compounds is thus incompatible with the Haldane scenario, contrary to the conclusions of Ref.~\onlinecite{fujihala2018} where direct evaluation of the coupling strength $\mathfrak J_{\rm 1D}$ was not attempted, and the gap size was not scaled against the relevant exchange coupling. We also note that magnetic ordering temperatures of $T_N/\mathfrak J_{\rm 1D}\simeq 0.87$ (A = Na) and 0.77 (A = K) are comparable to the coupling strength and would require a sizable interchain coupling $\mathfrak J_{\perp}/\mathfrak J_{\rm 1D}\simeq 0.3$~\cite{yasuda2005} or a large single-ion anisotropy to explain the formation of long-range magnetic order within any spin-chain scenario.}

The single-ion anisotropy is central to the 2D scenario too, because it facilitates long-range magnetic order even in the absence of any interplane coupling. This anisotropy should be also responsible for the opening of an excitation gap, which we illustrate by calculating magnetization curve using fitted values of $\mathfrak J$ and $\mathfrak D$ for Na$_2$Cu$_3$O(SO$_4)_3$ (Table~\ref{tab:inter}) and applying the field perpendicular to the easy direction. The spin-flop transition at 2.5\,T is indeed comparable to the weak kink observed in the experimental $M(H)$ around 3\,T (Fig.~\ref{fig:properties}f). 

\resub{Another important experimental observation is that the excitation gap of 0.6\,meV has been observed at 1.5\,K only. At 3\,K, this gap is already closed~\cite{furrer2018}, even though thermal energy is more than twice smaller than the gap size. This behavior would be unexpected in a purely Haldane system, but appears naturally in the 2D scenario, where the gap opens in the magnetically ordered state and vanishes upon approaching $T_N$. Temperature dependence of the gap suggests that this excitation gap is not of Haldane origin. }


\section{Summary and Outlook}
We revised here the magnetic models of the compounds A$_2$Cu$_3$O(SO$_4)_3$ represented by the fedotovite (A = K) and puninite (A = Na) minerals containing Cu$_6$ hexamer units. At low temperatures, each of these units develops the $S=1$ ground state and shows weak interactions to the neighboring hexamers. These interactions do not run along the crystallographic $b$ direction, as proposed previously. Our comprehensive analysis of the experimental data, as well as \textit{ab initio} modeling, suggest that all members of the series are better described by the $S=1$ square-lattice interaction topology. \cred{This puts Haldane physics into question, but highlights the importance of anisotropy of the $S=1$ macrospins.} 

\resub{Our results explain temperature dependence of the excitation gap as arising from magnetic anisotropy in the long-range-ordered state and not from the Haldane physics. This situation is not uncommon also in spin-1 chain materials, where weak interchain interactions and/or single-ion anisotropies may cause long-range magnetic order and affect low-energy excitations. Detailed neutron-scattering studies revealed that even in this case one may expect signatures of Haldane physics at higher energies, where spinon excitations were observed~\cite{zaliznyak2001,kenzelmann2001}. This may not be the case in A$_2$Cu$_3$O(SO$_4)_3$, though, because the interaction topology of these compounds is clearly 2D, whereas N\'eel temperatures are correspondingly higher and comparable to the energy scale of magnetic couplings $\mathfrak J$. Our revised microscopic magnetic model suggests more conventional magnon excitations that are typical for long-range-ordered antiferromagnets. This prediction would be interesting to verify once large enough single crystals become available. }

Magnetic interactions in A$_2$Cu$_3$O(SO$_4)_3$ show two distinct energy scales. Interactions within the hexamer are mostly determined by the Cu--O--Cu bond angles. On the other hand, interactions between the hexamers are controlled by the SO$_4$ tetrahedra that determine the admixture of second-neighbor oxygen atoms to the magnetic orbital of Cu$^{2+}$ (Fig.~\ref{fig:wannier}). The abundance of Cu-based sulphate minerals formed in volcanic fumaroles with highly oxidizing conditions~\cite{vergasova2016} can lead to different types of magnetic networks~\cite{inosov2018} tunable by chemical substitutions that control the orientation of the SO$_4$ tetrahedra and, therefore, superexchange pathways. Recent advances in synthetic procedures, which mimic natural geological processes on volcanoes~\cite{kovrugin2015,siidra2020}, make such compounds feasible in the lab and fully amenable to a detailed low-temperature characterization.

\acknowledgments
This work was supported by the Embassy of France in Russia, the French Government, and managed by the Agency Campus France. The Fonds Europ\'eens de D\'eveloppement R\'egional, CNRS, R\'egion Hauts-de-France, Contrat Plan Etat R\'egion, and Minist\'ere de l’Education Nationale de l’Enseignement Supe\'rieur et de la Recherche are acknowledged for funding the X-ray diffractometers and PPMS system. Claire Minaud and the Insitut Chevreul are also aknowledged for their experimental contribution. This work was partially carried out under the framework of the LOVE-ME project supported by the ANR (Grant ANR ANR-16-CE08-0023). AT was supported by the Federal Ministry for Education and Research through the Sofja Kovalevskaya Award of Alexander von Humboldt Foundation. O.S. was financially supported by the Russian Foundation for Basic Research, grant no. 19-05-00413.

%

\end{document}